\documentclass[twocolumn, trackchanges]{aastex63}

\usepackage{hyperref}
\usepackage{verbatim}
\graphicspath{{./}{figures/}}
\received{2020 July 11}
\revised{2020 August 30}
\accepted{2020 September 8}
\submitjournal{ApJ}
\usepackage{amsmath}
\usepackage{amsfonts}
\shorttitle{Redback PSR J2339$-$0533 }
\shortauthors{Kandel et al.}
\defcitealias{romani2011orbit}{RS11} 	
\defcitealias{Kandel_2020}{KR20}

\begin{document}

\title{Heated Poles on the Companion of Redback PSR J2339$-$0533 }

\correspondingauthor{D. Kandel}
\email{dkandel@stanford.edu}

\author[0000-0002-5402-3107]{D. Kandel}
\affil{Department  of  Physics,  Stanford  University,  Stanford,  CA, 94305, USA}

\author[0000-0001-6711-3286]{Roger W. Romani}
\affil{Department  of  Physics,  Stanford  University,  Stanford,  CA, 94305, USA}
\author[0000-0003-3460-0103]{Alexei V. Filippenko}
\affil{Department of Astronomy, University of California, Berkeley, CA, 94720, USA}
\affil{Miller Senior Fellow, Miller Institute for Basic Research in Science, University of California, Berkeley, CA 94720, USA}
\author{Thomas G. Brink}
\affil{Department of Astronomy, University of California, Berkeley, CA, 94720, USA}
\author{WeiKang Zheng}
\affil{Department of Astronomy, University of California, Berkeley, CA, 94720, USA}

\begin{abstract}
We analyze photometry and spectra of the ``redback" millisecond pulsar binary J2339$-$0533. These observations include new measurements from Keck and GROND, as well as archival measurements from the OISTER, WIYN, SOAR, and HET telescopes. The parameters derived from GROND, our primary photometric data, describe well the rest of the datasets, raising our confidence in our fitted binary properties. Our fit requires hot-spots (likely magnetic poles) on the surface of the companion star, and we see evidence that these spots move over the 8\,yr span of our photometry. The derived binary inclination $i = 69.3^\circ\pm 2.3^\circ$, together with the center-of-mass velocity (from the radial-velocity fits) $K_{\rm C} = 347.0\pm 3.7\,$ $\mathrm{km\,s}^{-1}$, give a fairly typical neutron star mass of $1.47\pm 0.09\,M_\odot$. 
\end{abstract}


\section{Introduction} \label{sec:intro}

Using optical imaging, \citet{romani2011orbit} and \citet{kong2012discovery} discovered a binary system with $P_B=0.193$\,d coincident with one of the brightest unidentified {\it Fermi}/LAT sources, inferring that it was the tidally locked, heated companion of a millisecond pulsar (MSP). Radio observations \citep{ray2014discovery, ray2020radio} found a 2.9\,ms pulsar at this position, which was generally obscured by a particularly powerful companion wind, but occasionally visible at 820\,MHz. The connection with the gamma-ray source was confirmed via gamma-ray pulsations \citep{pletsch2015gamma}; the source is an $\dot{E} = 2.32\times 10^{34}\,\mathrm{erg\,s}^{-1}$ ``redback"-type MSP with a low-mass main-sequence companion. This and the extreme spectroscopic variation (from mid $\sim 3500$\,K M-class spectra at minimum brightness to $\sim 7000$\,K F-class at maximum brightness) found by \citet[][hereafter RS11]{romani2011orbit} show that the companion is very strongly heated.

The initial photometry (\citetalias{romani2011orbit}; \citealt{yatsu2015multi}) sufficed to demonstrate this strong heating, but did not allow a detailed fit for the orbital parameters. Here we report on a new analysis of precision photometry, which shows highly significant asymmetries in the orbital light curves. Such asymmetry has been observed in other heated companions (e.g., \citealt{stappers2001intrinsic};  \citealt{schroeder2014observations}), and it has been suggested that these distortions may arise from asymmetric heating from the system's intrabinary shock \citep[IBS;][]{2016ApJ...828....7R} or from hot-spots on the companion magnetic poles created by precipitating IBS particles \citep{2017ApJ...845...42S}. Recently, \citet[][herafter \citetalias{Kandel_2020}]{Kandel_2020} have described a model in which global winds may advect the direct pulsar heating, also producing light-curve distortions. Each of these predicts somewhat different heating patterns. Our photometry, which includes four epochs over eight years, also indicates that the distortions are not constant. We find that a hot-spot which shifts location can reproduce these light curves, with consistent (and constant) geometric parameters for the binary. We combine this geometric information with a reanalysis of the \citetalias{romani2011orbit} Hobby-Eberly Telescope (HET) spectroscopic data to infer the neutron star mass as $1.47\pm 0.09\,M_\odot$. We conclude with a discussion of the nature of the hot-spot asymmetry and recommendations for future observations that seek to measure the mass of such binaries. 

\begin{figure*}[t!!]
    \centering
    \includegraphics[scale=0.45]{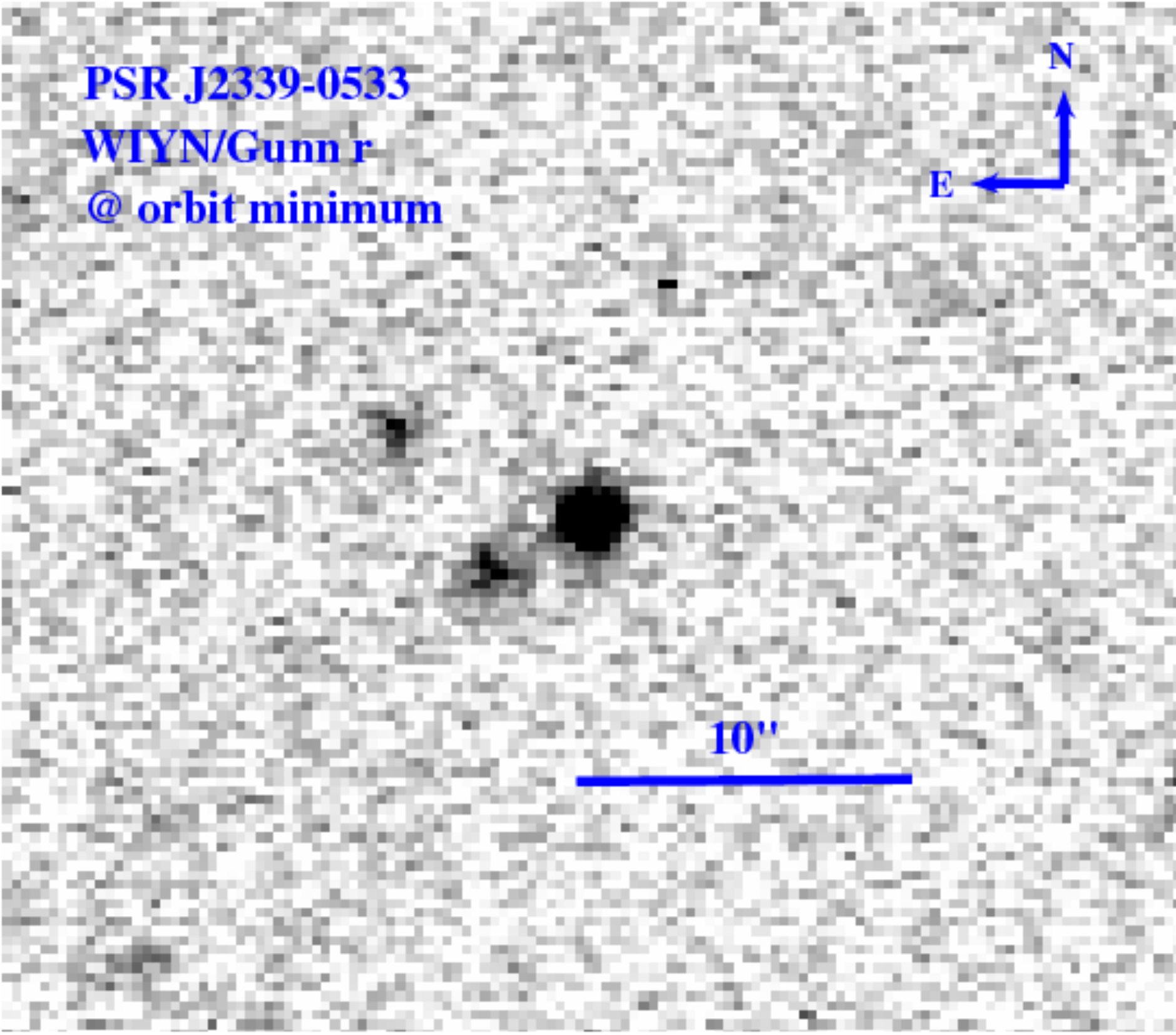}\hspace{0cm}
    \includegraphics[scale=0.40]{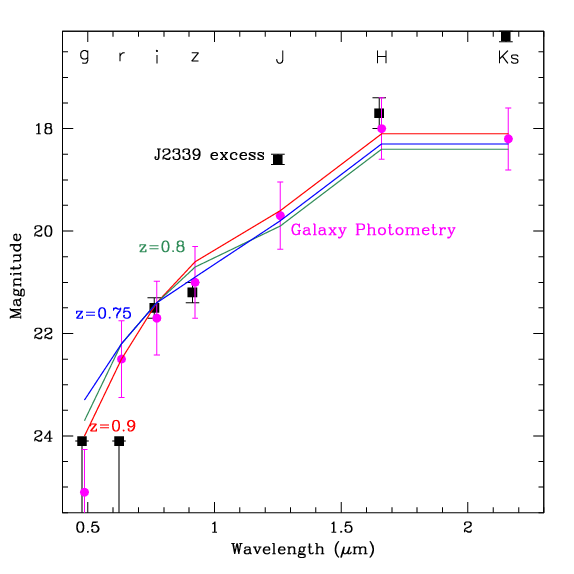}
    \caption{Left: WIYN Gunn $r$ image near orbital minimum, showing the pulsar companion and the contaminating galaxy $\sim 3^{\prime\prime}$ SE.
    Right: The SED of the contaminating source (magenta points), compared with the excess to the best-fit GROND model at minimum (black points). SEDs for elliptical galaxies at redshift $z=0.75$ (cyan line), $z=0.8$ (green line), and $z=0.9$ (red line) are shown for comparison.}
    \label{fig:PSR_im}
\end{figure*}

\section{Observations}

Our principal new photometric dataset is derived from an analysis of archival $grizJHK$ photometry of PSR J2339$-$0533 (hereafter J2339) taken on Sep. 11 and 14, 2017 (UT dates are used throughout his paper) with the GROND imager on the ESO 2.2\,m telescope (Program 099.A-9014). With $150\times 145$\,s for the optical exposures over the two observing sessions, the data covered 1.89 orbits. We downloaded the image frames and associated bias, flat, and dark frames. After standard IRAF processing and combination of the subexposures in the infrared (IR) frames, we extracted aperture photometry at the pulsar position measured near orbital maximum brightness. The instrumental magnitudes were calibrated against Sloan Digital Sky Survey (SDSS) measurements of field stars in the optical and against Two-Micron All-Sky Survey (2MASS) stars in the near-IR. Unfortunately, with the limited GROND field of view, only a handful of calibration stars were available. The seeing during these observations was poor and variable, with full width at half-maximum intensity (FWHM) $1.5-3.8''$, and the airmass was as large as $\sim 2.6$, leading to large apertures and low signal-to-noise ratio (S/N) detections near orbital minimum brightness.  Nevertheless, the photometry was quite stable, with useful detections throughout the orbit. We also extracted the IR-channel data, calibrating against a single nearby 2MASS star. We find that the $J$-band light curve is of good quality, and $H$ shows the heating effect, but the combination of limited S/N, large background uncertainties, and low $T_{\rm eff}$ sensitivity made the $K_s$ GROND data nearly useless. 

This photometry can be compared with more limited data taken at other epochs. First, optical imaging of J2339 was obtained at the WIYN 3.5\,m telescope with the MiniMo imager on Sep. 27-28, 2011 ($5\times240$\,s + $2\times180$\,s in Gunn $g$, $9\times120$\,s + 240\,s in Gunn $r$, $7\times120$\,s + 300\,s in Gunn $i$). We also examined $BVRI$ photometry from Sep. 22 to Oct. 7, 2011, taken by the OISTER collaboration \citep{yatsu2015multi} and kindly shared by those authors. Next, photometry at the SOAR 4.2\,m telescope using the GHTS in direct imaging mode on August 12, 2013, was collected: $3\times180$\,s + 300\,s in SDSS $u^\prime$, $4\times120$\,s + 300\,s in SDSS $g^\prime$, $4\times120$\,s + 60\,s in SDSS $r^\prime$, $4\times120$\,s + 300\,s in SDSS $i^\prime$, and $4\times600$\,s in H$\alpha$. For all SOAR and WIYN frames, standard CCD reductions were made and the fluxes were calibrated to SDSS stars in the field using $u^\prime$, $g^\prime$, $r^\prime$ (for Gunn $r$ and H$\alpha$), and $i^\prime$ magnitudes. For the WIYN data, the seeing image quality was good ($0.67^{\prime\prime}$ FWHM), and we can see (Fig. \ref{fig:PSR_im}, left) that a faint extended source lies near the pulsar counterpart.

\begin{figure*}[t!]
    \centering
    \hspace*{-1mm}\includegraphics[scale=0.45]{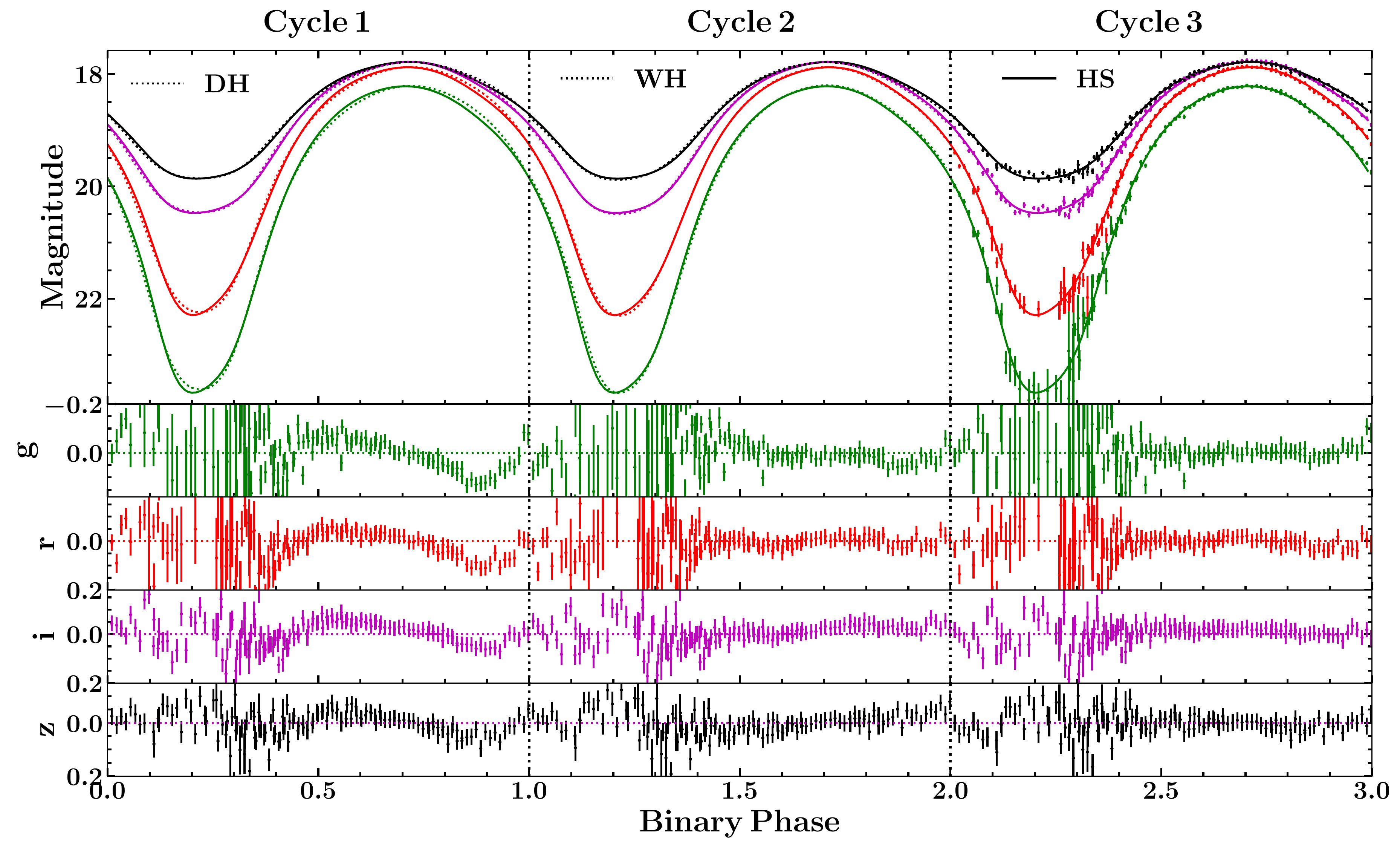}
    \caption{Light curves ({\it griz}) of J2339. Three periods are plotted with $\phi_B= 0$ at pulsar TASC (ascending node).  The solid curves in all three cycles show the best-fit hot-spot model (Table \ref{tab:hs}, row 3).  The dotted curve of cycle 1 shows the DH model with an arbitrary phase shift (fits without a phase shift are completely unacceptable). The dotted curve of cycle 2 illustrates the wind model.  Lower panels show fit residuals from the three models, with the hot-spot model (cycle 3) having the smallest residuals.}
    \label{fig:fit_LC}
\end{figure*}

Most recently, on Oct. 28, 2019, we obtained eight 600\,s exposures with the Keck-I 10\,m telescope plus the Low Resolution Imaging Spectrometer \citep[LRIS;][]{Oke95} in long-slit mode, using the 5600\,\AA\, dichroic, the 400\,line\,mm$^{-1}$ (blazed at 4000\,\AA) blue grism, and the 400\,line\,mm$^{-1}$ (blazed at 8500\,\AA) red grating, covering the binary orbital minimum brightness. This gave us spectra in the approximate range 3300--10,500\,\AA, with dispersions of 0.63\,\AA\,pixel$^{-1}$ (blue) and 1.2\,\AA\,pixel$^{-1}$ (red). The atmospheric dispersion corrector (ADC) allowed us to not have the slit aligned along the parallactic angle \citep{Fil82}, instead rotating the $1^{\prime\prime}$-wide slit so that we could simultaneously observe a nearby brighter G0 star. This enabled us to monitor the system throughput and to detect small shifts in the wavelength solution between frames. In addition, since this comparison (``comp") star has known and stable SDSS magnitudes, we are able to integrate the extracted pulsar and comp-star spectra over the appropriate wavelength ranges to get accurate relative photometry. This gives the pulsar magnitudes in broad-band filters, up to a possible small grey shift (due to different companion and comparison star slit losses) across the eight exposures. These spectra were followed by two $g/I$ image pairs, which served to check comp-star placement and stability. Thus, we have ten Keck photometric measurements in $g$ and $i/I$ and eight in other filters.

Of course, the spectral velocity  information is also directly useful, and we augment the new Keck spectra with a reanalysis of 48 600\,s exposures taken with the HET/LRS as described by \citetalias{romani2011orbit} to extract a new radial-velocity curve of the companion.

Although severely blended with the pulsar counterpart in the low-resolution GROND data, we have measured the spectral energy distribution (SED) of the nearby extended source and find it consistent with a galaxy at redshift $z \approx 0.8$--0.9 (Fig. \ref{fig:PSR_im}, right). The source is comparable in flux to J2339 at minimum brightness, and with the large apertures needed for the GROND data, it produces substantial contamination in the redder bands.

\section{Photometric Fitting}

Using a recent recomputation of the gamma-ray ephemeris \citep{an2020orbital} which provides excellent pulse-phase aligned timing throughout the {\it Fermi} mission, including the epochs of all observations described in this paper, we determine the binary phase from the barycentered time of the exposure midpoints. The GROND dataset is densely sampled with nearly uniform phase coverage, so we fit these data to best constrain the binary parameters.

The GROND $griz$ light curves (Fig. \ref{fig:fit_LC}) are well sampled and quite smooth.  First, the optical maximum brightness is shifted significantly later in phase than pulsar superior conjunction.
Any model which does not account for this is completely unacceptable. Accordingly, our minimal direct heating (DH) model must include an arbitrary (not physically justified) phase shift $\Delta\phi$.  Also, the light curve shows significant asymmetry, with excess emission on the leading side, especially in the bluer colors. This a clear sign of heat redistribution from the subpulsar point. 

We also fit for an extra background flux in each band, since the large-aperture GROND extractions ($3.2^{\prime\prime}$ in the optical, $5^{\prime\prime}$ in the IR) guarantee contamination by the nearby galaxy. The best-fit contamination fluxes do follow the red galaxy spectrum in the optical (Fig. \ref{fig:PSR_im}). The IR fluxes are somewhat larger; this may be due to the larger photometric aperture, but may also implicate a red nonphotospheric contribution from J2339 itself. The best-fit parameters of this DH model are given in Table \ref{tab:fit_J2339}. With a large $\chi^2$ per degree of freedom (DoF) of 2.51, it is unable to explain the asymmetric light curve, as can be seen in the fit residuals of Figure \ref{fig:fit_LC}.

\begin{deluxetable}{lccc}[t!!]
\tabletypesize{\footnotesize}
\tablewidth{0pt}

 \tablecaption{2017 GROND light-curve fit results for J2339\label{tab:fit_J2339}}

 \tablehead{
 \colhead{Parameters} & \colhead{DH + Phase Shift} & \colhead{WH} & \colhead{HS}}
 \startdata 
 $i\,(\mathrm{ deg})$ & $58.4^{+0.7}_{-0.7}$ & $55.9^{+0.5}_{-0.5}$ & $69.3^{+2.3}_{-2.3}$\\
  $f_c$ & $0.95^{+0.01}_{-0.01}$  & $0.97^{+0.01}_{-0.01}$ & $0.97^{+0.02}_{-0.02}$\\
 $L_{\mathrm{P}}/(10^{34}$\,erg\,s$^{-1}$) & $2.26^{+0.05}_{-0.05}$  & $2.53^{+0.04}_{-0.04}$ & $1.48^{+0.03}_{-0.03}$ \\
 $T_N$\,(K) & $3183^{+28}_{-29}$ &$3126^{+28}_{-30}$ & $3307^{+64}_{-90}$\\
 $d$\,(kpc) & $1.92^{+0.01}_{-0.01}$  & $1.97^{+0.01}_{-0.01}$ & $1.87^{+0.01}_{-0.01}$\\
 $\epsilon$ & ...& $0.31^{+0.006}_{-0.004}$ & ...\\
 $\theta_c\, (\mathrm{deg})$ &...&$55.3^{+2.0}_{-1.6}$ & ...\\
 $\Delta\phi$ & $-0.030^{+0.001}_{-0.004}$ & ... & ...\\
 $\chi^2/\mathrm{DoF}$ &$1388/553$  & $877/552$ & $671/550$ 
 \enddata
\end{deluxetable}

One way to produce light-curve phase shifts and asymmetry is via a global circulation, as in the model developed by \citetalias{Kandel_2020}, where an equatorial wind redistributes heat from the subpulsar point. This wind is characterized by $\epsilon\equiv \tau_{\rm{rad}} \omega_{\rm{adv}}$, the ratio of radiation time to advection time at the equator (prograde for $\epsilon > 0)$. Such winds have been inferred for several ``hot Jupiters" \citep{2011ApJ...726...82C}.
The models generally have flows reversing direction at mid-latitudes; in our model, we fit for $\theta_c$, the angle from the equator at which the flow (with the same $\epsilon$) changes sign. Hydrodynamical models of such flows can show more complex patterns, and \citet{2020arXiv200606552V} have recently developed a similar model also incorporating heat diffusion, but our simple prescription captures the bulk heat redistribution with sufficient freedom to use in model fits. The fit with this wind-heating (WH) model implies a super-rotating equatorial wind resulting in the overall phase shift of the light-curve maxima by $\Delta\phi_{\rm{max}}\approx -0.03$. The $\chi^2$ decrease of this model is large, with strong statistical preference over the DH model. 

However, there is good reason to expect that the low-mass stellar companions of redbacks are significantly magnetized so that the companion field can channel energetic pulsar particles to heat its surface at magnetic caps formed by the field foot-points \citep{2017ApJ...845...42S}. Thus, we also fit with a single hot-spot (HS) model having a simple Gaussian excess on the companion surface, characterized by amplitude $\mathcal{A}_{\rm hs}$, radial size $r_{\rm hs}$, and angular position $\theta_{\rm hs}, \phi_{\rm hs}$. 
The binary parameters for this model are listed in Table \ref{tab:fit_J2339}. The hot-spot parameters (Table \ref{tab:hs}) indicate a substantial (32\%) temperature excess in a large-radius ($47^\circ$) pole. This pole is located in the companion's ``southern" hemisphere (across the equator from Earth's line of sight) and leads the subpulsar point near $L_1$. The fit is superior to that of both the DH and WH models. The $\chi^2/$DoF approaches 1 and the fit residuals reduce greatly (Fig. \ref{fig:fit_LC}, panel 3), showing that the model reproduces the observed asymmetry quite well. One should note that $i$ is substantially higher (and $L_p$ is lower) for the HS model than for the DH and WH models. The other binary parameters are similar.

\begin{figure}
\centering
\hspace*{-5mm}\includegraphics[scale=0.40]{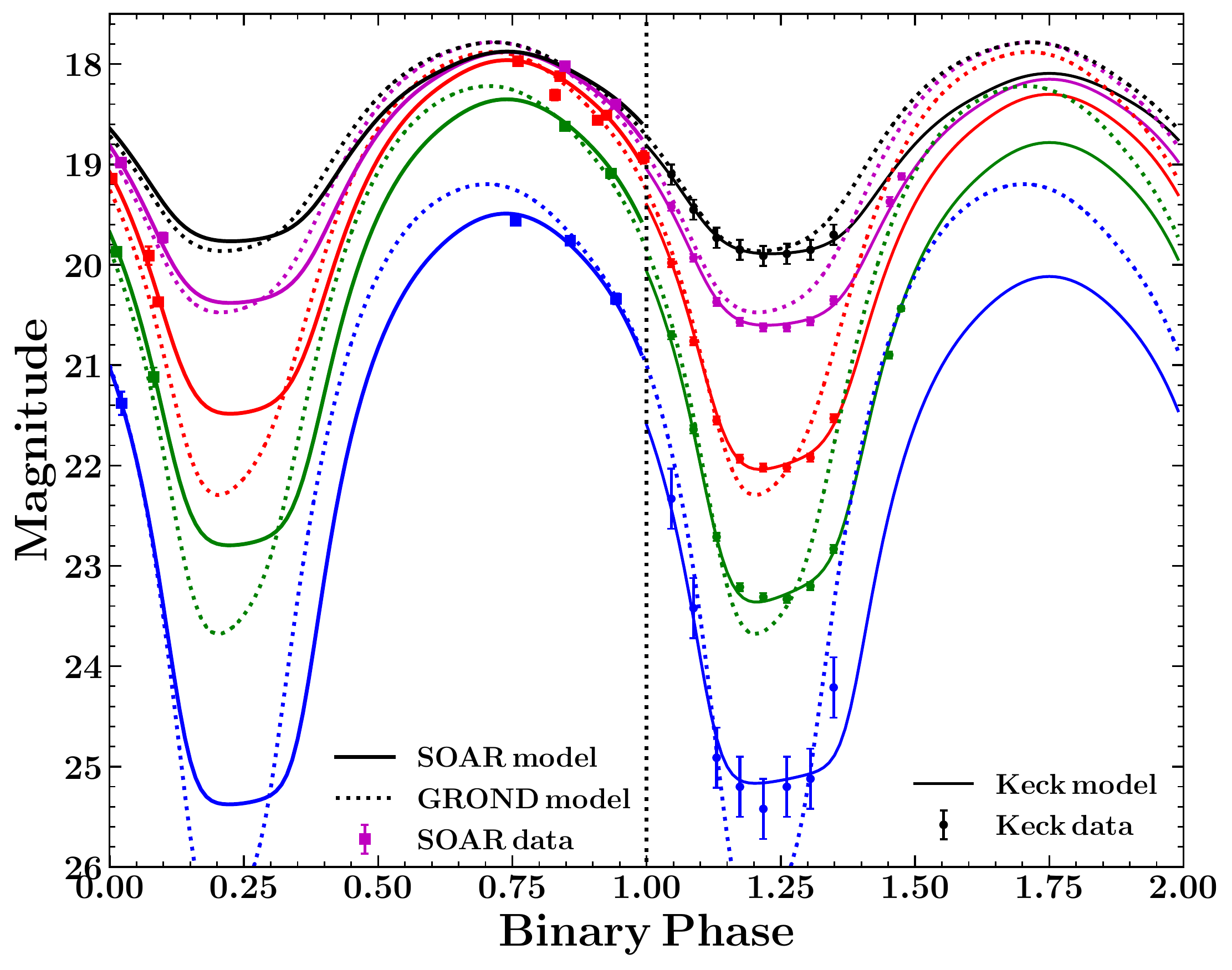}
\caption{Left: 2013 {\it ugri} SOAR photometric observations, compared with the best-fit HS model (solid curves, Table \ref{tab:hs}, row 2).  Right: 2019 {\it ugriz} Keck photometry and best-fit HS model (solid curves, Table 2, row 4). For comparison, the dotted {\it griz} curves in both panels are the best-fit HS model for the GROND epoch solid curves of Fig. \ref{fig:fit_LC} (Table \ref{tab:hs}, row 3).}
    \label{fig:keck}
\end{figure}

\begin{figure}
\centering
\hspace*{-5mm}\includegraphics[scale=0.40]{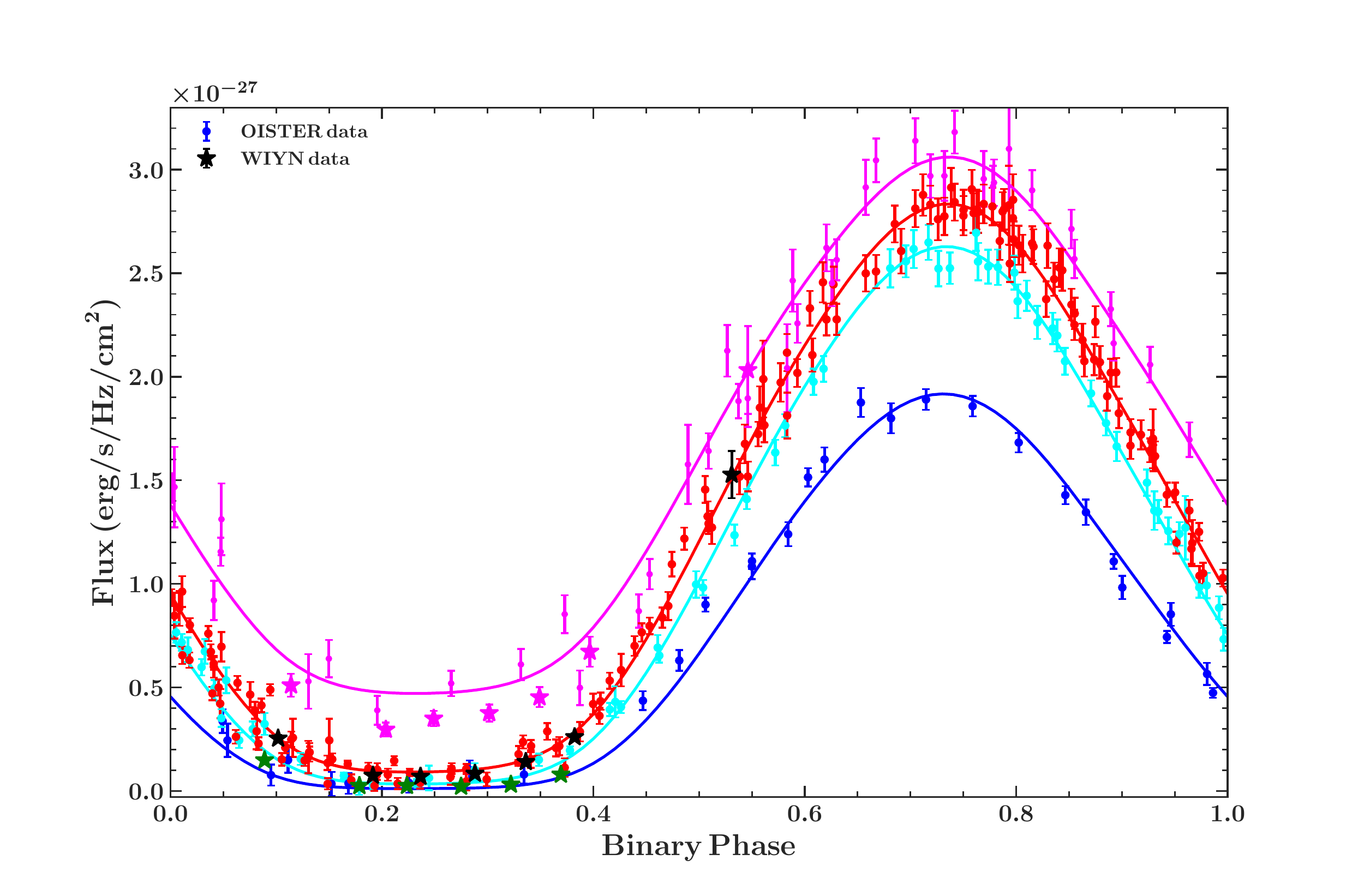}
\caption{{\it BVRI} OISTER photometry from \citet{yatsu2015multi} and best-fit HS model (solid curves, Table \ref{tab:hs}, row 1). The WIYN {\it g} (green), {\it r} (black), and {\it i} (magenta) points at similar epoch are overlaid.}
\label{fig:oister}
\end{figure}

\section{Shifting hot-spots}

Armed with the binary parameters determined by the fit to the GROND data, we can check consistency with the partial light curves provided by our other datasets, which 
span eight years. First, the sparse WIYN 2011 data show a minimum appreciably brighter than the GROND model curve, with the minimum closer to $\phi=0.25$ than the GROND data. Near this epoch (September 22 -- October 7, 2011), observations were made using the Optical and Infrared Synergetic Telescopes for Education and Research (OISTER). Originally presented by \citet{yatsu2015multi}, this dataset has good phase coverage and shows a phase shift. We have elected to fit these datasets together for the 2011 epoch; if fit separately, both indicate a hot-spot at similar $\theta_{\rm hs}$ and $\phi_{\rm hs}$. Next, 2013 SOAR photometry covered only maximum brightness, but also show a peak slightly later in phase than for GROND. Perhaps the best comparison, though, is with the 2019 Keck data. With the spectral points we have multicolor coverage of the orbital minimum, plus a few late $g/I$ points. This minimum is distinctly bluer than in the GROND data, with a flat minimum well centered on $\phi=0.25$.

Of course, orbital parameters should not change over the eight years. Instead, we posit that the heating pattern has changed; in particular, the position of the magnetic pole (hot-spot) may shift and the pulsar illumination may change. Thus, we fit each of the four epochs with the orbital parameters of Table \ref{tab:fit_J2339} fixed, but the hot-spot parameters free. The results are in Table \ref{tab:hs} and the resulting light curves are shown in Figure \ref{fig:keck} and Figure \ref{fig:oister}. With limited phase coverage the parameters are not always well constrained, but three interesting features appear: (i) all spots are large, (ii) all are located in the southern hemisphere ($\phi_{\rm hs}<0$, across the equator from the Earth line-of-sight), and (iii) spots on the ``day" (pulsar) side ($\theta_{\rm hs} < 90^\circ$) are more strongly heated (larger fractional temperature increase $\mathcal{A}_{\rm hs}$) than the Keck example, which is on the back ``night" side. If the hot-spots are heated by precipitation of IBS particles ducted to the companion, as in the model of \citet{2017ApJ...845...42S}, then fewer particles are captured by field lines extending away from the pulsar, so the weaker heating fit for the Keck example is natural.

As for all redbacks, this companion is a low-mass star, fully convective in the core with a short spin period imposed by tidal locking, so we expect a strong $\alpha-\Omega$ dynamo as well as a strong and frequently refreshed magnetic field. So, magnetic pole hot-spots are natural and changes in the dipole axis are plausible. Of course, the regenerated field may assume a random orientation under each regeneration -- this is a nominal conclusion from our fit spot locations. However, it is intriguing that all four epochs are consistent with $\phi_{\rm hs} \approx -60^\circ$ to $-80^\circ$; in this case, we might infer that the magnetic axis is relatively stable, but that the differing $\theta_{\rm hs}$ could represent a drifting interior dipole. Such a motion may explain the shifting light curve of redback PSR J1723$-$2837 interpreted as asynchronous companion rotation by \citet{van2016active}. Certainly, our data cannot distinguish these possibilities, but a future sensitive multicolor campaign can probe this feature.

\begin{deluxetable*}{lcccccc}
\tabletypesize{\footnotesize}
\tablewidth{0pt}
 \tablecaption{Best-fit hot-spot parameters for different epochs (in chronological order). \label{tab:hs}}
 
 \tablehead{
 \colhead{Dataset} & \colhead{Obs MJD} & \colhead{$\theta_{\rm{hs}}\,(\rm{deg})$} & \colhead{$\phi_{\rm{hs}}\,(\rm{deg})$} & \colhead{$\mathcal{A}_{\rm{hs}}$} & \colhead{$r_{\rm{hs}}\,(\rm{deg})$} & \colhead{Ref.}}
 \startdata
 WIYN + OISTER & 55826 - 55841& $65.2\pm 2.4$ & $-79.3\pm 2.2$ & $0.54\pm 0.10$& $31.3\pm 6.3$ & Fig. \ref{fig:oister}\\ 
 SOAR & 56516& $85.0\pm 8.0$ & $-80.1\pm 5.7$ & $0.40\pm 0.20$& $40.0\pm 11.5$ & Fig. \ref{fig:keck}\\ 
 GROND & 58007 - 58010& $70.1\pm 1.1$ & $-53.3^{+5.1}_{-4.6}$ &$0.43^{+0.05}_{-0.03}$ &$33.5^{+4.0}_{-3.4}$ & Fig. \ref{fig:fit_LC} \\ 
 Keck & 58784 & $124.5\pm 17.2$ & $-59.0^{+26.0}_{-13.8}$ & $0.10^{+0.05}_{-0.01}$ & $45.8\pm 11.7$ & Fig. \ref{fig:keck}\\ 
\enddata
\end{deluxetable*}

If the companion magnetic field is dominated by a dipole, we might expect particles ducted to both hemispheres, but with lower efficiency toward the back side. So, we fit with opposing spots having identical sizes, but free heating amplitudes $\mathcal{A}_{\rm hs}$ for the two hemispheres. For the GROND epoch, we refit the full model; reassuringly, all fit binary parameters are within the uncertainty of single-spot fit values. We find $\mathcal{A}_1/\mathcal{A}_2 \approx 10$. For the Keck epoch the flux ratio is relatively unconstrained, $1.6 \lesssim\mathcal{A}_1/\mathcal{A}_2\lesssim 7.5$, but the brighter (northern) hemisphere pole is poorly constrained mostly because of the lack of data around the light-curve maxima.
It will be interesting if future intensive studies can test the expectation that heated poles will have the largest power when closest to the companion nose. 

\section{Spectral Analysis}

\begin{figure}
    \centering
    \includegraphics[scale=0.4]{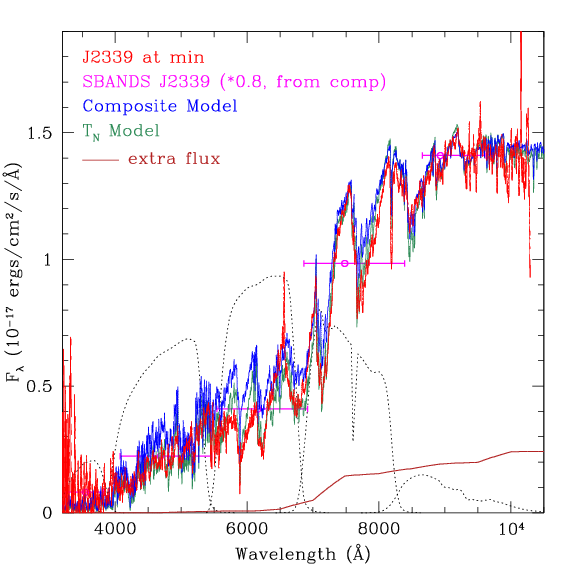}
    \caption{J2339 Keck spectrum, averaged over the four exposures at minimum brightness (red). This is compared with the model composite spectrum (blue) as well as a single night-side base temperature model (green). The model spectra include extra flux toward the red: $i=2.7\pm 0.5\,\mu$Jy and $z=5.1\pm 1.5\,\mu$Jy. The average colors synthesized from the spectrum with the {\tt sbands} routine for the SDSS bands (faint dotted curves) are shown by the magenta points.}
    \label{fig:minspec}
\end{figure}

We can compare the Keck spectroscopy with the photometric fit model. Figure \ref{fig:minspec} shows the Keck flux averaged over the four spectra at the light-curve minimum. For comparison, we show the composite model spectrum (blue) and a single-temperature $T_N$ model averaged over the same four Keck nighttime phases. The composite spectrum is computed using the {\tt ICARUS} code \citep{breton2012koi} for a model with reduced $\gamma$-ray heating, a shifted hot-spot (Table \ref{tab:hs}, row 4), 
and excess IR flux attributed to the background galaxy (Fig. \ref{fig:PSR_im}, left). The general agreement is reasonable, with an M3--M1-class spectrum, but the composite model is too blue. The companion also has a strong H$\alpha$ line, with weaker H$\beta$ visible in some spectra.

\begin{figure}
    \centering
    \includegraphics[scale=0.4]{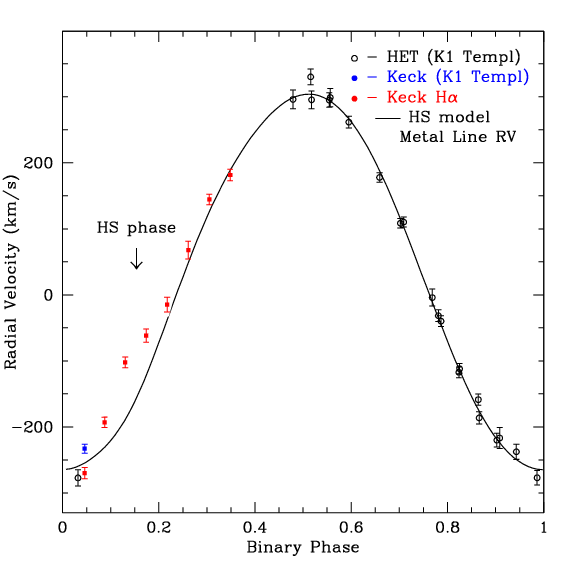}
    \caption{J2339 radial-velocity measurements, from metal lines (K1 template) and H$\alpha$ emission. The best-fit curve, employing the Keck-epoch host-spot illumination, is shown.}
    \label{fig:rv}
\end{figure}

Since they are dominated by molecular bands, the Keck spectra at minimum brightness  do not provide good radial velocities. The first spectrum at $\phi_B=0.046$, however, contains sufficiently strong metal lines that we can fit for a radial velocity using a K1-star template. In addition, the H$\alpha$ line provides good velocities for all spectra. We have also compared with the HET spectra of \citetalias{romani2011orbit}, remeasuring the radial velocities by fitting with a K1 template while excluding 100\,\AA\ ranges around the Balmer absorption features that dominate near optical maximum. No evidence for H$\alpha$ emission is seen in the lower-S/N, lower-resolution HET data. Retaining only the HET points with strong cross-correlation peaks (from the day phases), we obtain the radial-velocity curve of Figure \ref{fig:rv}. The HET spectra seem to have a wavelength offset, which introduced a substantial $\Gamma \approx -80\,\mathrm{km\,s}^{-1}$ in \citetalias{romani2011orbit}; we have chosen to match the Keck velocity solution for the K1 fits, which result in a small positive $\Gamma$.

As emphasized by \citet{linares2018peering} and discussed by \citetalias{Kandel_2020}, different line species have varying temperature sensitivities and so are differently distributed across the face of the companion. By fitting with K1 templates (and excluding the Balmer-line wavelengths), we are most sensitive to the metal lines.

\section{System Modeling and Discussion}

For a given heating model, the radial-velocity data can be used to infer the companion center-of-mass (CoM) motion. Adopting the GROND-epoch light-curve model (Table \ref{tab:fit_J2339}), we can compute the equivalent width (EW)-weighted radial velocity at each orbital phase, for a given line species. Here, since we use a K1 template to measure the radial velocities, we are most sensitive to the common metal absorption lines. Using a set of archival standard dwarf spectra, we have computed the temperature dependence of the EWs of the strongest Fe, Mg~I, and Na~I optical lines. Averaging, a simple power-law fit gives ${\rm EW}(T) = \left(3.57/T_{3000}\right)^{2.71}$. With this prescription, we compute the metal-line radial-velocity curve for a given model and can fit to the spectroscopic data to determine the CoM radial velocity $K$ and $\Gamma$. Although we do not perform a simultaneous fit with photometric data, the spectroscopic fits are marginalized over the geometrical parameters from the end of the  GROND photometric  Markov Chain Monte Carlo chains, sampling $2\sigma$ uncertainties.  Thus, the mass errors include all uncertainties in the model fitting, spectroscopic and photometric. 

The fit results are given in Table \ref{tab:vel_J2339} while the best-fit radial-velocity curve is shown Figure \ref{fig:rv}. For our base model (HS model) this gives a companion CoM velocity amplitude $K_{\rm C} = 347.0\pm 3.7$\,$\mathrm{km\,s}^{-1}$, a relatively modest neutron star mass of $1.47\pm 0.09\, M_\odot$, and a companion mass of $0.30\pm 0.02\, M_\odot$. Note that this companion mass is consistent with its observed spectral class and fit base temperature of $\sim 3300$\,K. 

\begin{deluxetable}{lccc}
\medskip
\tabletypesize{\footnotesize}
\tablewidth{0pt}

 \tablecaption{Radial-velocity fit results for J2339\label{tab:vel_J2339}}

 \tablehead{
 \colhead{Parameters} & \colhead{Phase Shift} & \colhead{WH} & \colhead{HS}}
 \startdata 
 $K_{\rm{C}} \,(\mathrm{km\,s}^{-1})$ & $351.3^{+3.7}_{-3.7}$ & $353.7^{+3.8}_{-3.7}$ & $347.0^{+3.9}_{-3.6}$\\
 $\Gamma\,(\mathrm{km\,s}^{-1})$ &  $22.8^{+2.1}_{-2.1}$ & $15.2^{+2.1}_{-2.1}$ & $17.7^{+2.1}_{-2.1}$\\
 $M_{\rm{NS}}\,(M_\odot)$ & $2.02^{+0.07}_{-0.07}$& $2.22^{+0.08}_{-0.08}$ & $1.47^{+0.09}_{-0.09}$\\
$M_{\rm{C}}\,(M_\odot)$ & $0.40^{+0.01}_{-0.01}$& $0.44^{+0.01}_{-0.01}$ & $0.30^{+0.02}_{-0.02}$\\
 $\chi^2/\mathrm{DoF}$ & $23/19$  & $22/19$ & $23/19$
 \enddata
\end{deluxetable}

One model mass uncertainty in our study is the heating pattern (which differs at different epochs) used for estimation of the CoM velocity. To quantify the uncertainty, we fit the CoM velocity using the heating pattern of the other three epochs. The resulting velocities are $351.1\pm 3.8\,\mathrm{km\,s}^{-1}$, $349.3\pm 3.7\,\mathrm{km\,s}^{-1}$, and $348.6\pm 3.7\,\mathrm{km\,s}^{-1}$ for Keck, WIYN + OISTER, and SOAR, respectively. Such differences lead to mass shifts of $\sim \pm 0.05\, M_\odot$. This shows that although the different heating models do imply small differences in the CoM radial-velocity amplitude $K_c$ and hence mass,  the largest differences arise from the different inclinations of the fit models.
The other (deprecated) heating models give substantially different masses. For the phase-shift or WH model, one would infer large ($\gtrsim 2.0\,M_\odot$) masses. Some systems may indeed have such large masses (see \citetalias{Kandel_2020}), but here the smaller mass HS solution is clearly statistically preferable. Additional support can be drawn from the fact that the high-mass models have $M_C > 0.4\,M_\odot$, inconsistent with the low $T_N$ of the light-curve fits.

The connection between the H$\alpha$ radial velocities and the underlying CoM velocity is unclear. Interestingly, the largest departures from the model radial-velocity curve are near the inferred hot-spot phase. However, the relative redshift of the emission line is a puzzle. If it represented outflow from the companion surface, a blueshift would be expected at these phases. Additional spectroscopy with good S/N might follow this line emission into the day side of the orbit, giving clues to its origin.

Note that here we have determined the radial velocities by adopting a cross-correlation fit dominated by metal lines and then using the model surface temperature distribution to map the EW-weighted radial velocity. A more complete analysis would be to generate model spectra for each phase and to cross-correlate these spectra with the data to find the radial-velocity shifts uniformly fit from all spectral features (using a range of species with different $T$ dependence). For objects such as J2339 with a large ($>2\times$) range in the surface temperature, this would be the best way to connect back- (night) side molecular band shifts with the day-side Balmer line velocities. We plan to pursue such an analysis in upcoming papers.

Our evidence for secular light-curve variations joins that for other redbacks. It seems that this is a quite common feature of these systems and we speculate that it is associated with time-varying magnetic dipoles on the active companion, heated by precipitating IBS particles. We encourage high-quality multiband light curves of these systems at few-month separations over several years to probe the physical origin of these variations.\\


We thank the anonymous referee for a very detailed and careful reading of the manuscript.
We are grateful for the excellent assistance of the staffs of the observatories where data were taken.
Some of the data presented herein were obtained at the W. M. Keck Observatory, which is operated as a scientific partnership among the California Institute of Technology, the University of California, and NASA; the observatory was made possible by the generous financial support of the W. M. Keck Foundation.   
D.K. and R.W.R. were supported in part by NASA grants 80NSSC17K0024 and 80NSSC17K0502.
A.V.F.'s group is grateful for generous financial assistance from the Christopher R. Redlich Fund, the TABASGO Foundation, and the Miller Institute for Basic Research in Science (UC Berkeley).

\bibliographystyle{aasjournal}
\bibliography{main}
\end{document}